\documentclass[12pt]{article}
\makeatletter
\@addtoreset{equation}{subsection}
\makeatother

\begin{document}
\begin{center}
{\Large \bf Cosmological Quantum Jump Dynamics} \\[0.5cm]
{\large\bf II. The Retrodictive Universe }\\[1.5cm]
  {\bf Vladimir S.~MASHKEVICH}\footnote {E-mail:
   Vladimir\_Mashkevich@qc.edu}
\\[1.4cm] {\it Physics Department
  \\ Queens College\\ The City University of New York\\
  65-30 Kissena Boulevard\\ Flushing, New York
  11367-1519} \\[1.4cm] \vskip 1cm

{\large \bf Abstract}
\end{center}

This paper is a continuation of the paper [1] and is dedicated to
the problem of the arrow of time. A deterministic past-directed
dynamics is constructed, which results in the retrodictive
universe. A future-directed dynamics of the latter is
indeterministic and reproduces standard probabilistic quantum
dynamics. The arrow of time is inherent in the retrodictive
universe as well as a future-directed increase of informational
entropy.

\newpage

\section*{Introduction}

One of the most ancient and difficult problems of physics is that
of the origin of the arrow of time, i.e., of the nature of the
difference between the past and the future. It is conventional to
search for a solution to this problem in dynamics, i.e., time
evolution of a state of a physical system. The solution may be
given by a dynamics which is asymmetric, or orientable in the
sense of the sequence of states.

A deterministic dynamics does not involve such an orientability.
Therefore it seems advisable to search for the solution in an
indeterministic dynamics. But, according to an established
opinion, in the standard indeterministic quantum dynamics there is
no arrow of time as well [2,3].

A crucial feature of an orientable dynamics is the prevalence of
retrodiction over prediction. An essential mathematical concept
inherent in indeterminism is that of randomness [4-6]. It is the
latter that seems to be the main impediment to the construction of
a dynamics with the above-mentioned feature.

The simplest way of constructing an orientable dynamics is to
construct one in which a past-directed time evolution would have
been deterministic. In doing this, the pivotal point is the
condition that a future-directed time evolution represent the
standard indeterministic, i.e., probabilistic quantum dynamics.

In the present paper, that idea is realized. We introduce a
jump-deciding mechanism for a two-state tangency vertex. The
mechanism consists of the Planck clock (one with the Planck time
period) and rules which decide a quantum jump by the reading of
the clock. On the basis of this mechanism, a deterministic
past-directed dynamics is constructed. The corresponding
future-directed dynamics turns out to be indeterministic. The
rules are selected in such a way that probabilities be standard
quantum ones.

The construction outlined above exhibits a retrodictive universe.
Its salient features are the following. A complete retrodiction
and a partial prediction are involved. It is impossible to
introduce initial conditions and to extend dynamics forward in
time; thus the retrodictive universe is constructed at once as a
whole---for all times from a maximal future to a maximal past. In
this connection we quote Weyl [7]: ``The objective world simply
is, it does not {\it happen}. Only to the gaze of my
consciousness, crawling upward along the life line of my body,
does a section of this world come to life as a fleeting image in
space which continuously changes in time.''

\section{The problems of the arrow of time and randomness}

\subsection{The problem of the time arrow}

We quote Roger Newton [8]: ``One of the greatest puzzles of
physics is the manifest discord between two facts: on one hand,
all the fundamental equations and laws of physics are
(essentially) invariant under time reversal;\ldots on the other
hand, we are all aware that at the macroscopic level, many
physical processes flow in one time direction only. The
unidirectional flow of time \ldots is one of the most obvious
features both of our consciousness and of the physical world,
\ldots which near the end of the nineteenth century presented
physics with one of the most profound challenges, and about which
there are, to this day, strong disagreements among physicists.''

There are five arrows of time [8]:

1. The delay between cause and effect.

2. The biological, or {\it cognitive} arrow.

3. The second law of thermodynamics.

4. The cosmological arrow (the expansion of the universe).

5. The direction of the time parameter used in physics.

The fifth arrow is present in any dynamics. The second and third
arrows are related by an informational aspect: information on the
past is greater than on the future, so that (informational)
entropy increases in time. It is this aspect that forms the basis
for our attacking the time arrow problem.

\subsection{The problem of randomness}

The mathematical concept of randomness is inherent in an
indeterministic, i.e., probabilistic dynamics. A physical problem
related to the concept is the impossibility of empirically
verifying or falsifying the randomness of results of a generic
quantum dynamical process. We quote Beltrami [4]: ``\ldots there
can be {\it no formal proof that a sufficiently long string is
random} \ldots In response to the persistent question `{\it Is it
random?}' the answer must now be `probably, but I'm not sure';
`probably' because most numbers are, in fact, random \ldots and
`not sure' because the randomness of long strings is essentially
undecidable.'' Thus the question of the randomness of quantum
dynamics cannot be, strictly speaking, decided empirically. So
there remains the possibility of choosing a decision which would
provide a resolution  of some theoretical problems.

\subsection{Time nonorientability of the standard
indeterministic quantum dynamics}

There exists no arrow of time in the standard indeterministic
quantum dynamics. This is exhibited by means of a fully
time-symmetric construction from which conventional quantum
mechanics may be derived [9] (see a detailed treatment in [3]).

Since the conventional indeterministic (probabilistic) dynamics is
based on the concept of randomness, it seems reasonable to try to
abandon the latter in constructing an orientable dynamics: we do
not see what else may be done.

\section{A deterministic past-directed quantum jump\\ dynamics}

\subsection{Tangency vertex}

Let us consider a tangency vertex [1], i.e., a confluence or/and
branch point of levels without crossing, in more detail. Assume
that without quantum jumps the Hamiltonian $H(t)$ is a $C^\infty$
operator-valued function of time, so that for a vertex
\begin{equation}\label{2.1.1}
\frac{d^n H^- _{\rm ver}}{dt^n}=\frac{d^nH^+_{\rm
ver}}{dt^n}=\frac{d^nH_{\rm ver}}{dt^n}, \quad n=0,1,2,\ldots
\end{equation}
holds. In an infinitesimal neighborhood of the vertex,
\begin{equation}\label{2.1.2}
H(t_{\rm ver}+\Delta t)=H[g(t_{\rm ver}+\Delta t)]= H[g(t_{\rm
ver})+\dot{g}(t_{\rm ver})\Delta t]
\end{equation}
and
\begin{equation}\label{2.1.3}
H_{\rm ver}(t_{\rm ver}\mp\Delta
t)=\sum_l^{1,n^\mp}\varepsilon_l(t_{\rm ver}\mp\Delta t)P_l(t_{\rm
ver}\mp\Delta t),\quad \Delta t\geq 0
\end{equation}

For any projector $E(t)$
\begin{equation}\label{2.1.4}
E\frac{dE}{dt}E=0
\end{equation}
is fulfilled, so that
\begin{equation}\label{2.1.5}
\frac{d^n}{dt^n}\left(E\frac{dE}{dt}E\right)=0,\quad
n=0,1,2,\ldots
\end{equation}
Hence it is easily seen that from
\begin{equation}\label{2.1.6}
\frac{d^mE^-}{dt^m}=\frac{d^mE^+}{dt^m} \qquad {\rm for}\quad
m=0,1,\ldots,n
\end{equation}
follows
\begin{equation}\label{2.1.7}
\left(E\frac{d^{n+1}E}{dt^{n+1}}E\right)^-=
\left(E\frac{d^{n+1}E}{dt^{n+1}}E\right)^+
\end{equation}
Now for a tangency vertex, we obtain by the method of [1]
\begin{equation}\label{2.1.8}
\frac{d^nP_{\rm tan}^-}{dt^n}=\frac{d^nP_{\rm
tan}^+}{dt^n}=\frac{d^nP_{\rm tan}}{dt^n}, \qquad n=0,1,2,\ldots
\end{equation}
where
\begin{equation}\label{2.1.9}
P^\mp_{\rm tan}=\sum_l P_l^\mp=P_{\tan}
\end{equation}
and
\begin{equation}\label{2.1.10}
\frac{d^n \varepsilon^-_l}{dt^n}=\frac{d^n
\varepsilon^+_{l'}}{dt^n}=\frac{d^n
\varepsilon_{\tan}}{dt^n},\qquad n=0,1,2,\ldots
\end{equation}
Thus, at a tangency vertex, there is contact of order $n=\infty$.

\subsection{A two-state jump-deciding mechanism in a past-directed\\ dynamics}

We assume that the degeneracy of levels occurs only at vertices,
so that there are only two-state vertices.  Therefore we consider
a two-state tangency vertex. For the latter, we introduce a
jump-deciding mechanism. The mechanism consists of a Planck clock
and rules deciding a jump by the reading of the clock. The Planck
clock is one with the Planck time $t_{\rm P}$ period. (There is no
other natural time interval.) We will measure time in units of
$t_{\rm P}$ so that the period of the clock is 1.

There are two states $i=1,2$ for $t<t_{\tan}$ and two states
$f=1,2$ for $t>t_{\tan}$. We consider transitions, i.e., quantum
jumps $i\rightarrow f$ in a future-directed dynamics and
transitions $f\rightarrow i$ in a past-directed dynamics. It is
the latter that has to be deterministic, so that let $f=a$ be an
actual state before a quantum jump $a\rightarrow i$. Quantum
probabilities are
\begin{equation}\label{2.2.1}
p(f|i)=p_{f\rightarrow i}=p_{i\rightarrow f}=|\langle
f|i\rangle|^2
\end{equation}
with
\begin{equation}\label{2.2.2}
\sum_f p(f|i)=\sum_i p(f|i)=1
\end{equation}
Let
\begin{equation}\label{2.2.3}
\epsilon_{i=1}<\varepsilon_{i=2}
\end{equation}

Introduce some $\bar{t}$,
\begin{equation}\label{2.2.4}
0\leq \bar{t}\leq 1
\end{equation}
A deterministic past-directed dynamics is defined by the following
jump-deciding rule for $a\rightarrow i$:
\begin{equation}\label{2.2.5}
{\rm if}\quad 0\leq t_{\rm tan}<\bar{t}\quad {\rm then}\quad
a\rightarrow 1,\qquad {\rm if}\quad \bar{t}\leq t_{\rm tan}<1\quad
{\rm then}\quad a\rightarrow 2
\end{equation}

\subsection{A future-directed dynamics}

In order to provide a correct probabilistic future-directed
dynamics, we introduce
\begin{equation}\label{2.3.1}
p_{\rm min/max}=({\rm min/max})_i\{p(a|i)\},\quad p_{\rm min}\leq
1/2\leq p_{\rm max}
\end{equation}
\begin{equation}\label{2.3.2}
t_1=\frac{p_{\rm min}}{1/2+p_{\rm min}},\quad
t_2=\frac{1/2}{1/2+p_{\rm min}},\quad t_1\leq t_2,\quad t_1+t_2=1
\end{equation}
and put
\begin{equation}\label{2.3.3}
\bar{t}=t_i \quad {\rm where}\quad p(a|i)=p_{\rm min}
\end{equation}

The rule (2.2.5) is equivalent to these:
\begin{equation}\label{2.3.4}
{\rm if} \quad 0\leq t_{\rm tan}<t_1\quad {\rm then}\quad
a\rightarrow 1,\qquad {\rm if}\quad t_2\leq t_{\rm tan}<1\quad
{\rm then}\quad a\rightarrow 2
\end{equation}
\begin{equation}\label{2.3.5}
{\rm if}\quad t_1\leq t_{\rm tan}<t_2\quad{\rm then}\quad
a\rightarrow i\quad {\rm where}\quad p(a|i)=p_{\rm max}>1/2\quad
({\rm here}\quad t_1<t_2)
\end{equation}

Now if an initial state is $i$, then, in view of (2.3.4),
\begin{equation}\label{2.3.6}
t_{i-1}\leq t_{\rm tan}<t_{i+1}
\end{equation}
where
\begin{equation}\label{2.3.7}
0=t_0\leq t_1\leq t_2\leq t_3=1
\end{equation}
and
\begin{equation}\label{2.3.8}
t_{i+1}-t_{i-1}=t_2
\end{equation}
Therefore we should have for probabilities
\begin{equation}\label{2.3.9}
p_{\rm min}=\frac{t_1/2}{t_2},\quad p_{\rm
max}=\frac{(t_2-t_1)+t_1/2}{t_2}=\frac{t_2-t_1/2}{t_2}
\end{equation}
since $t_1\leq t_{\rm tan}<t_2$ implies $p(a|i)=p_{\rm max}$,
whereas if $t_{\rm tan}<t_1$ or $t_{\rm tan}\geq t_2$ then
$p(a|i)=p_{\rm min}$ and $p(a|i)=p_{\rm max}$ are equiprobable.
The choice (2.3.3) provides (2.3.9).

\subsection{The Planck clock and two fundamental theories}

For the sake of generality, we might choose any time period
$t_{\rm per}$ rather than the Planck time $t_{\rm P}$. We have put
\begin{equation}\label{2.4.1}
t_{\rm per}=t_{\rm P}
\end{equation}
for lack of any other natural time interval. There is an added
reason for incorporating the Planck clock into quantum jump
dynamics, i.e., putting (2.4.1).

The Planck time
\begin{equation}\label{2.4.2}
t_{\rm P}=\left(\frac{\hbar G}{c^5}\right)^{1/2}=5.3906\times
10^{-44}{\rm s}
\end{equation}
may be introduced as a natural unit for all physical quantities.
Put
\begin{equation}\label{2.4.3}
c=1
\end{equation}
then
\begin{equation}\label{2.4.4}
\hbar G=t_{\rm P}^2
\end{equation}
Next, there are two systems of units:
\begin{equation}\label{2.4.5}
\hbar=1,\quad G=t_{\rm P}^2
\end{equation}
(natural units) and
\begin{equation}\label{2.4.6}
G=1,\quad \hbar=t_{\rm P}^2
\end{equation}
(``geometrized units'' [10]). In the natural units, general
relativity involves $t_{\rm P}$: it appears in the Einstein
equation, but the standard quantum theory does not involve $t_{\rm
P}$: it does not appear in the Schr$\ddot{\rm o}$dinger equation
and quantum probabilities. By contrast, in the geometrized units,
it is quantum theory rather than general relativity that involves
$t_{\rm P}$.

The Planck clock incorporates $t_{\rm P}$ into quantum theory via
quantum jump dynamics. Now, in the natural units, general
relativity and quantum theory enjoy equal rights with respect to
$t_{\rm P}$, which links those fundamental theories in addition to
the construction introduced in [1].

\section{The retrodictive universe and its salient features}

\subsection{The retrodictive universe}

The construction accomplished above exhibits a universe which is
naturally called retrodictive. Let us consider its salient
features.

\subsection{A complete retrodiction and a partial prediction}

Since the past-directed dynamics is deterministic, in the
retrodictive universe there exists a complete retrodiction, to
which the universe owes its name.

An informational aspect of retrodiction is this: A complete
information on the universe's states in the past reduces to and
may be obtained from the information contained in the description
of a present state and in the laws of the past-directed dynamics.

The future-directed dynamics is indeterministic and reproduces the
dynamics of the standard probabilistic quantum theory.
Notwithstanding this fact, there exists a partial prediction
provided by (2.3.5): if
\begin{equation}\label{3.2.1}
t_1\leq t_{\rm tan}<t_2
\end{equation}
then in the future-directed dynamics the jump $i\rightarrow f$
happens to the $f$ for which
\begin{equation}\label{3.2.2}
p(f|i)=p_{\rm max}>1/2
\end{equation}
But this partial prediction does not violate the probabilistic
relations (2.3.9).

\subsection{The impossibility of introducing initial conditions}

The past-directed dynamics constructed backwards in time starting
from some state at $t=t_{\rm initial}$, $-\infty<t_{\rm
initial}<\infty $, cannot be extended forward in time: we may run
into a situation where
\begin{equation}\label{3.3.1}
i=1,\quad t_2\leq t_{\rm tan}<1\qquad {\rm or}\quad i=2,\quad
0\leq t_{\rm tan}<t_1
\end{equation}
which is inconsistent with the rule (2.3.4). Thus it is impossible
to construct a future-directed dynamics starting from some initial
conditions. It is only a past-directed dynamics that may be
constructed in the case of the retrodictive universe.

\subsection{The entirety of the retrodictive universe}

The impossibility of constructing a future-directed dynamics
starting from some initial conditions given at some $t_{\rm
initial}$, i.e., solving the Cauchy problem, implies the entirety
of the retrodictive universe: the latter is determined at once for
all times from a maximal future to a maximal past. The universe
exists but does not evolve in time [7]. A seeming evolution is the
result of the indeterministic character of the future-directed
dynamics.

This conclusion cracks the problem of initial conditions for the
universe: there are none.

\subsection{The arrow of time}

The arrow of time is inherent in the retrodictive universe as well
as a future-directed increase of informational entropy.

\section*{Acknowledgments}

I would like to thank Alex A. Lisyansky for support and Stefan V.
Mashkevich for helpful discussions.

\end{document}